\documentclass[12pt,a4paper]{article}
\usepackage{latexsym,amsmath,amssymb}
\date{}
\numberwithin{equation}{section}

\begin{document}
\title{Van der Waerden calculus with commuting spinor variables and the Hilbert-Krein structure of the superspace}
\author{Florin Constantinescu\\ Fachbereich Mathematik \\ Johann Wolfgang Goethe-Universit\"at Frankfurt\\ Robert-Mayer-Strasse 10\\ D 60054
Frankfurt am Main, Germany}
\maketitle

\begin{abstract}
Working with anticommuting Weyl (or Majorana) spinors in the frame of the van der Waerden calculus is standard in supersymmetry. The natural framework of rigorous supersymmetric quantum field theory makes use of operator-valued superdistributions defined on supersymmetric test functions. This makes necessary a van der Waerden calculus in which the Grassmann variables anticommute but the fermionic components are commutative instead of being anticommutative. We work out such a calculus in view of applications to the rigorous conceptual problems of supersymmetric quantum field theory.
\end{abstract}

\section{Introduction}
It was recently shown that the $N=1$ superspace carries an inherent indefinite Krein structure \cite{C}. It can be turned as usual into a Hilbert space structure \cite{Stro} providing us with what we have called the standard Hilbert-Krein structure of the $N=1$-superspace \cite{C}. The corresponding Hilbert space of supersymmetric functions is the counterpart of the classical $L^2 $-space induced by a Lorentz invariant measure. The main point of our construction was the SUBSTRACTION of the transversal sector $P_c +P_a -P_T $ instead of its addition to the chiral/antichiral sectors suggested by the decomposition $P_c +P_a +P_T =1$ where $P_i ,i=c,a,T $ are the formal chiral, antichiral and transversal projection operators \cite{WB}. We believe that the supersymmetric Hilbert-Krein structure is needed for futher developpements in rigorous supersymmetric quantum field theory (for some first applications see \cite{C}). \\ 
A general supersymetric function will be denoted by

\begin{gather}\nonumber 
X(z)=X(x,\theta ,\bar \theta )= \\ \nonumber
=f(x)+\theta \varphi (x) +\bar \theta \bar \chi (x) +\theta ^2m(x)+\bar \theta^2n(x)+ \\ 
\theta \sigma^l\bar \theta v_l(x)+\theta^2\bar \theta \bar \lambda(x)+\bar \theta^2\theta \psi (x)+ \theta^2 \bar \theta^2d(x)
\end{gather}
where the coefficients are functions of $x$ in Minkowski space of certain regularity which will be specified below. For the moment let us suppose that they belong to the Schwartz space $S$ of infinitely differentiable functions with faster than polynomial decrease at infinity. It is usual in supersymmetries to work with fields of the form (1.1) for which the spinorial coefficients $\varphi ,\bar \chi ,\bar \lambda ,\psi $ are anticommuting with the Grassmann variables $\theta_\alpha ,\bar \theta_\alpha ,\alpha =1,2 $ and between themselves. This is part of the common wisdom telling us that the fermions have to anticommute. Whereas this perfecty fits functional integral methods, it may be obstructive when trying to rigorously define supersymmetric quantum fields as operator valued superdistributions in the same way as the usual quantum fields are defined to be operator valued distributions \cite{StreW}. Indeed superdistributions have to be defined as linear functionals over (test) superfunctions of the type (1.1) with commuting numerical coefficient functions $f,\varphi ,\bar \chi ,\bar \lambda ,\psi ,m,n,v_l ,d $. Besides the fact that properly speaking numerical functions cannot be anticommuting, the commutativity of the coefficients in (1.1) makes possible the interpretation of the quantum supersymmetric fields in terms of their components as the collection of the "ordinary" quantum fields, bosons and fermions, defined as operator valued distributions. \\
Certainly supersymmetric fields have to carry noncommutative spinor coefficients. Our disscussion above concerns only the test functions to which they have to be applied. On the other hand the work with anticommuting Weyl or Majorana spinors is very much simplified by the van der Waerden calculus which is standard in supersymmetry \cite{WB,Striv,Wei}. Using the rules of computation of this calculus we have uncovered in \cite{C} the Hilbert-Krein definite/indefinite structure of the $N=1$ superspace. 
Certainly one has the well justified feeling that, at least in \cite{C}, adopting anticommuting spinor components is just an oportunistic convention which simplifies the computations by using existing formulas in the vast supersymmetric literature, whereas the Hilbert-Krein structure itself should not depend on this convention. In other words we should be able to derive the standard Hilbert-Krein structure of $N=1$ supersymmetry using commutative spinor components too. The aim of the present paper is to provide the necessary details proving the above mentioned statement. The main business is to turn the noncommutative van der Waerden calculus into a commutative one; more precisely into a mixed one in which the Grassmann variables remain anticommuting whereas the spinor components are by now communing objects, and to show that this fact has no effects on the net results. In this way the bona-fide superdistributional framework of quantum superfields \cite{C}, which has to cope with test functions of form (1.1) with commuting spinor components, is completely worked out. 

\section{Commutative van der Waerden calculus}

The van der Waerden calculus \cite{vdW,PR,WB,Striv} turns spinor matrix algebra into a spinor tensor calculus common for vectors and tensors. It is used in supersymmetry together wirth an overall convention of anticommutativity of the Grassmann variables and spinor components. We follow here the notations and conventions of \cite{WB,Striv}. The main tools are  antisymmetric "metric tensors" $(\epsilon^{\alpha \beta }),(\epsilon_{\alpha \beta }),\epsilon_{21}=\epsilon^{12}=1,\epsilon_{12}=\epsilon^{21}=-1  $ and the Pauli $\bar \sigma $ matrix

\begin{gather}
\bar \sigma^{l\bar \alpha \alpha }=\epsilon^{\bar \alpha \bar \beta }\epsilon^{\alpha \beta }\sigma^{\beta \bar \beta }
\end{gather}
where $\sigma_{\alpha ,\bar \alpha }^l,l=0,1,2,3 $ are the usual Pauli matrices. Note the standard index positions for $\sigma $ and $\bar \sigma $ which make contact to the matrix interpretation: up for $\bar \sigma $, down for $\sigma $. Spinors $ \psi $ with upper and lower indices are related through the $\epsilon $-tensor:

\begin{gather}
\psi^\alpha =\epsilon^{\alpha \beta }\psi_\beta , \psi_\alpha =\epsilon_{\alpha \beta }\psi^\beta
\end{gather} 
One defines the anticommuting products

\begin{gather}
\psi \chi =\psi^\alpha \chi_\alpha =-\psi_\alpha \chi^\alpha =\chi^\alpha \psi_\alpha =\chi \psi \\
\bar \psi \bar \chi =\bar \psi^{\dot \alpha }\bar {\chi_{\dot \alpha }}=-\bar {\psi_{\dot \alpha }}\bar {\chi^{\dot \alpha }}=\bar {\chi^\alpha }\bar {\psi_\alpha }=\bar \chi \bar \psi
\end{gather} 
The conjugation $(\chi_\alpha )*=\bar \chi_{\dot \alpha }, (\chi^\alpha )*=\bar \chi^{\dot \alpha } $ reverses the order of spinor components: 

\begin{gather}
(\chi \psi )*=\overline {\chi \psi }=\overline {\chi^\alpha \psi_\alpha }=\bar \psi_{\dot \alpha }\bar \chi^{\dot \alpha }=\bar \psi \bar \chi =\bar \chi \bar \psi =\overline {\psi \chi }=(\psi \chi )*
\end{gather}
In view of the above relations we use a bar to denote complex conjugation (not the bar on $ \sigma $!). We have the usual list of rules in the (anticommuting) spinor algebra given for instance in Appendix A and Appendix B of \cite{WB}. Note that our conventions and notations coincide entirely with those of \cite{Striv} and with those of \cite{WB} up to the sign of the Pauli matrix $\sigma^0 $ which is minus one in \cite{WB} but one in \cite{Striv} and in the present paper. Concerning the differential and integral calculus in the Grassmann variables we follow usual conventions too (see for instance \cite{Striv}).
In particular the complex conjugation of the Grassmann derivative is defined to be $ (\partial_\alpha )*=\overline{\partial_\alpha }=-\bar \partial_{\dot \alpha }, \partial_\alpha =\frac{\partial }{\partial \theta^\alpha },\bar \partial_{\dot \alpha }=\frac{\partial }{\partial \bar \theta^{\dot \alpha }} $. \\
As long as we work in physics with supersymmetric fields the (anticommutative) van der Waerden calculus described above is very useful. But if we want to apply them to supersymmetric functions of the form (1.1) with numerical coefficient functions the rules have to change. Suppose that the spinor components are assumed to commute between themsels and with the Grassmann variables. Up to this modification we retain all conventions above. In particular the anticommuting property of the Grassmann variables remains unchanged. We start this paper by giving the corresponding counterpart of the rules in Appendices A and B of \cite{WB} for the commutative spinor algebra. They are:

\begin{gather}
\psi \chi =\psi^\alpha \chi_\alpha =-\psi_\alpha \chi^\alpha =-\chi^\alpha \psi_\alpha =-\chi \psi \\
\bar \psi \bar \chi =\bar \psi^{\dot \alpha }\bar {\chi_{\dot \alpha }}=-\bar {\psi_{\dot \alpha }}\bar {\chi^{\dot \alpha }}=-\bar {\chi^{\dot \alpha }}\bar {\psi_{\dot \alpha }}=-\bar \chi \bar \psi \\
\overline {\chi \psi }=\overline {\chi^\alpha \psi_\alpha }=\bar \chi^{\dot \alpha }\bar \psi_{\dot \alpha }=\bar \psi_{\dot \alpha }\bar \chi^{\dot \alpha }=\bar \psi \bar \chi =-\bar \chi \bar \psi =-\overline {\psi \chi } \\
(\theta \phi )(\theta \psi )=\frac{1}{2}(\phi \psi )\theta^2 \\
(\bar \theta \bar \phi )(\bar \theta \bar \psi )=\frac{1}{2}(\bar \phi \bar \psi )\bar \theta^2 \\
\chi \sigma^n \bar \psi =\bar \psi \bar \sigma^n \chi \\
\overline {\chi \sigma^n \bar \psi }=-\psi \sigma^n \bar \chi
\end{gather} 
We see that there are many sign discrepances in comparisson to the anticommuting convention. But some combined equations which are frequently used, as for instance $\overline {\chi \sigma^n \bar \psi }=-\bar \chi \bar\sigma^n \psi $, remain unchanged. Certainly there will be some consequences. In particular the conjugate function $\bar X(z)$ is different:

\begin{gather} \nonumber
\bar X=\bar X(x,\theta ,\bar \theta )=\\ \nonumber
=\bar f(x)-\theta \chi (x) -\bar \theta \bar \varphi (x) +\theta ^2 \bar n(x)+\bar \theta^2 \bar m(x)+ \\ 
+\theta \sigma^l \bar \theta \bar v_l(x)-\theta^2 \bar \theta \bar \psi(x)-\bar \theta^2\theta \lambda (x)+ \theta^2 \bar \theta^2 \bar d(x)
\end{gather}
as compared with the anticommuting case \cite{WB}

\begin{gather} \nonumber
\bar X=\bar X(x,\theta ,\bar \theta )=\\ \nonumber
=\bar f(x)+\theta \chi (x) +\bar \theta \bar \varphi (x) +\theta ^2 \bar n(x)+\bar \theta^2 \bar m(x)+ \\ 
+\theta \sigma^l \bar \theta \bar v_l(x)+\theta^2 \bar \theta \bar \psi(x)+\bar \theta^2\theta \lambda (x)+ \theta^2 \bar \theta^2 \bar d(x)
\end{gather}
Let us introduce the supersymmetric covariant (and invariant \cite{WB,Striv}) derivatives $D,\bar D$ with spinorial components $D_{\alpha },D^{\alpha
},\bar D_{\dot \alpha },\bar D^{\dot \alpha }$
which independently of conventions are given by

\begin{gather}  
D_{\alpha }=\partial_{\alpha } +i\sigma_{\alpha \dot \alpha }^l\bar
\theta^{\dot \alpha }\partial _l \\ 
D^{\alpha }=\epsilon ^{\alpha \beta }D_{\beta }=-\partial ^{\alpha }+i\sigma^{l\alpha }_{\dot \alpha }\bar \theta^{\dot \alpha }\partial _l  \\
\bar D_{\dot \alpha }=-\bar {\partial }_{\dot \alpha } -i\theta ^{\alpha }\sigma _{\alpha \dot \alpha }^l\partial _l \\ 
\bar D^{\dot \alpha }=\epsilon ^{\dot \alpha \dot \beta }\bar D_{\dot
  \beta }=\bar {\partial }^{\dot \alpha }-i\theta ^{\alpha }\sigma
_{\alpha }^{l\dot \alpha }\partial _l   
\end{gather}
We accept on the way notations like
\[\epsilon^{\alpha \beta }\sigma^l _{\beta \dot \alpha }=\sigma^{l\alpha
}_{\dot \alpha } \] etc., but in the end we will come back as usual to the canonical index positions $\sigma^l =(\sigma_{\alpha \dot \alpha }^l ),\bar \sigma^l =(\bar \sigma^{l\dot \alpha \alpha })$.\\
Note that $D_{\alpha }$ does not contain the variable $\theta $ and $\bar D ^{\dot \alpha } $ does not contain the variable $\bar \theta $ such that we can write at the operatorial level:

\begin{gather}
D^2=D^{\alpha }D_{\alpha }=-(\partial ^{\alpha }\partial _{\alpha }-2i\partial _{\alpha \dot \alpha }\bar \theta ^{\dot \alpha } \partial ^{\alpha }  +\bar \theta^2 \square ) \\   
\bar D^2=\bar D_{\dot \alpha }\bar D^{\dot \alpha }=-(\bar {\partial } _{\dot \alpha }\bar {\partial }^{\dot \alpha }+2i \theta ^{\alpha }\partial _{\alpha \dot \alpha }\bar \partial ^{ \dot \alpha } +\theta^2\ \square )    
\end{gather}
where 
\[\square =\eta ^{lm}\partial _l \partial _m \]
is the d'Alembertian, $\eta $ is the Minkowski metric tensor and
\[ \partial _{\alpha \dot \alpha }=\sigma _{\alpha \dot \alpha }^l \partial _l \]
Note that the complex conjugate of the formal operators $ D_\alpha ,D^\alpha ,D^2 $ are the formal operators $ \bar D_{\dot \alpha },  \bar D^{\dot \alpha }, \bar D^2 $ respectively. We will prove that in the Hilbert space to be still defined, $ \bar D^2 $ is also the operator adjoint of $ D^2 $. 
Note in passing that applied to a general supersymmetric function the "operator" $D_\beta $ produces:

\begin{gather} \nonumber 
D_\beta X=\varphi_\beta +\theta^\alpha (2m\epsilon_{\beta \alpha })+\bar \theta_{\dot \alpha }(-v_\beta ^{\dot \alpha }-i\sigma_\beta^{l\dot \alpha}\partial_l f)+ \\ \nonumber
+\bar \theta^2 (\psi_{\beta } -\frac{i}{2}\sigma_{\beta \dot \beta }^l \partial_l \bar \chi^{\dot \beta })+\theta^\alpha \bar \theta^{\dot \alpha }(2\epsilon_{\alpha \beta }\bar \lambda_{\dot \alpha }-i\sigma_{\beta \dot \alpha }\partial_l \varphi_\alpha )+  \\
+\theta^2 \bar \theta_{\dot \alpha }(-i\sigma_\beta^{l\dot \alpha }\partial_l m)+\bar \theta^2 \theta^\alpha (2\epsilon_{\beta \alpha }d+\frac{i}{2}\sigma_\beta^{l\dot \beta }\partial_l v_{\alpha \dot\beta })-\frac{i}{2}\theta^2 \bar \theta^2 \sigma_{\beta \dot \beta }^l \partial_l \bar \lambda^{\dot \beta }
\end{gather}
i.e. if we start with anticommuting spinor components in $X$ we do not preserve this property. Indeed the coefficient of the $\theta $-contribution in (2.21) is not an anticommuting spinor. In fact the covariant derivatives $D_\beta ,\bar D_{\dot \beta } ;\beta , \dot \beta =1,2 $ turn out to be nasty, nonmathematical objects in the anticommuting case but perfect Hilbert space operators in the commuting case (see later). Fortunately in physics we use only combinations of covariant derivatives which preserve spinor anticommutativity saving the situation as this is well known. If we start (as we definitely do in this paper) with supersymmetric functions with commuting spinor components some problems of mathematical consistency, as for instance the, admitingly harmless, problem above posed by $D_\beta ,\bar D_{\dot \beta } ;\beta ,\dot \beta =1,2 $ do not appear anyway . \\
We make use of the operators \cite{WB,Striv}:

\begin{gather}
c=\bar D^2D^2, a=D^2\bar D^2, T=D^{\alpha }\bar D^2 D_{\alpha }=\bar
D_{\dot \alpha }D^2 \bar D^{\dot \alpha }=-8\square +\frac {1}{2}(c+a) 
\end{gather}
which are used to construct formal projections

\begin{gather}
P_c=\frac {1}{16\square }c,P_a=\frac {1}{16\square }a, P_T=-\frac
{1}{8\square }T 
\end{gather}
on chiral, antichiral and transversal supersymmetric functions. These operators are, at least for the time being, formal because they contain the d'Alembertian in the denominator. Problems with the d'Alembertian in the denominator will be explained later. Chiral, antichiral and transversal funtions are linear subspaces of general supersymmetric functions (1.1) which are defined by the conditions \cite{WB,Striv}
\[ \bar D^{\dot \alpha }X=0, \dot \alpha =1,2,;  D^{ \alpha }X=0, \alpha =1,2; D^2 X=\bar D^2 X=0 \]
respectively. It can be proven that these relations are equivalent to the relations

\begin{gather}
cX=16\square X, aX=16\square X ,TX=-8\square X
\end{gather} 
respectively. \\
We have formally 
\[ P_i^2=P_i, P_iP_j=0,\quad i\ne j;i,j=c,a,T \]
and $P_c +P_a +P_T =1 $. Accordingly each supersymmetric function can be formally decomposed into a sum of a chiral, antichiral and transversal contributions. From a rigorous point of view this statement may be wrong and we have to reconsider it later in this paper because of the problems with the d'Alembertian in the denominators (it is true for the massive and wrong for the massless case). 
Let us specify the coefficient functions in (1.1) for the chiral,
antichiral and transversal supersymmetric functions. \\
For the chiral case $X_c$ we have:

\begin{gather} \nonumber 
\bar \chi =\psi =n=0 , v_l=\partial_l (if)=i\partial_l f ,
  \\ \bar \lambda =-\frac {i}{2}\partial_l \varphi \sigma^l
   =-\frac{i}{2}\bar \sigma^l \partial _l \varphi , 
  d=\frac{1}{4}\square f 
\end{gather}
Here $ f,\varphi $ and $ m $ are arbitrary functions. \\
For the antichiral $X_a$ case:

\begin{gather} \nonumber
\varphi =\bar \lambda =
m=0,  v_l=\partial_l (-if)=-i\partial_l f , \\ \psi =\frac{i}{2}\sigma^l
\partial_l \bar \chi =\frac{i}{2}\partial _l \bar \chi \bar \sigma^l , d=\frac{1}{4}\square f 
\end{gather}
Here $f,\bar \chi $ and $ n $ are arbitrary functions. \\
For the transversal case $X_T$ \cite{Striv}:

\begin{gather} \nonumber
m=n=0, \partial_l v^l =0, \\ \nonumber
\bar \lambda =\frac {i}{2}\partial_l \varphi
\sigma^l =\frac{i}{2}\bar \sigma^l \partial_l \varphi ,\psi
=-\frac{i}{2}\sigma^l \partial_l \bar \chi  =-\frac{i}{2} \partial_l \bar
\chi \bar \sigma^l \\
d=-\frac{1}{4}\square f 
\end{gather}
Here $f,\varphi ,\bar \chi $ are arbitrary functions and $v$ satisfies $\partial_l v^l =0 $. The above formulas are very similar to the usual one in the supersymmetric literature \cite{WB,Striv,C} but not quite identical to them. The reason is our commutative convention which in particular modifies the sign in front of $\bar \sigma $ (see relation (2.11 )).\\
Note that in the massless case there is an overlap between chiral/antichiral and transversal sectors. It appears here as a trouble but it will turn out to be even a chance. From (2.25)-(2.27) it follows that a function $X$ belongs to this overlap if

\begin{gather}\nonumber
X(z)=f(x)+\theta \varphi (x)+\bar \theta \bar \chi (x)\pm i\theta \sigma_l \bar \theta \partial_l f(x)
\end{gather}
with
\[\partial_l \varphi \sigma^l =\sigma^l \partial_l \bar \chi =0,\quad \square f=0 \]
Clearly in the massive case there are no nontrivial on-shell functions in the overlap. The overlap plays a major role for the massless vector field \cite{C}. In the Hilbert space approach to follow the overlap will contain only factorizable vectors of zero length. \\
We need also the following relations:

\begin{gather}\nonumber
\bar D^2 X=-4n+\theta (-4\psi -2i\sigma^l \partial_l \bar \chi )
+\theta^2 (-4d -2i\partial _l v^l-\square f)+\theta \sigma^l \bar \theta
(-4i\partial _l n)+\\ \nonumber
 +\theta^2 \bar \theta (2i\bar \sigma^l \partial _l \psi  -\square \bar \chi ) +\theta^2 \bar \theta^2 (-\square n) \\ \nonumber
D^2 X=-4m+\bar \theta (-4\bar \lambda +2i\bar \sigma^l \partial_l \varphi )+\bar \theta ^2(-4d +2i\partial _l v^l-\square f)+\theta \sigma^l \bar \theta (4i\partial _l m)+\\ \nonumber
+\bar \theta^2 \theta (-2i\sigma^l\partial_l \bar \lambda -\square \varphi )+\theta^2 \bar \theta^2 (-\square m) 
\end{gather}
These relations can be rewriten in a form which makes clear that $\bar D^2 X ,D^2X $ are chiral and antichiral respectively

\begin{gather}\nonumber
\bar D^2 X=-4n+\theta (-4\psi -2i\sigma^l \partial_l \bar \chi ) +\theta^2 (-4d -2i\partial _l v^l-\square f)+\theta \sigma^l \bar \theta (-4i\partial _l n)+\\ 
+\theta^2 \bar \theta (-\frac{1}{2}i\bar \sigma^l \partial_l )(-4\psi -2i\sigma^m \partial_m \bar \chi ) +\theta^2 \bar \theta^2 (-\square n) \\ \nonumber
D^2 X=-4m+\bar \theta (-4\bar \lambda +2i\bar \sigma^l \partial_l \varphi )+\bar \theta ^2(-4d +2i\partial _l v^l-\square f)+\theta \sigma^l \bar \theta (4i\partial _l m)+\\ 
+\bar \theta^2 \theta (\frac{1}{2}i\sigma ^l \partial_l )(-4\bar \lambda +2i\bar \sigma^m \partial_m \varphi ) +\theta^2 \bar \theta^2 (-\square m) 
\end{gather}
These expressions have to be compared with the corresponding expressions in the anticommuting case \cite{C}:

\begin{gather}\nonumber
\bar D^2 X=-4n+\theta (-4\psi -2i\sigma^l \partial_l \bar \chi ) +\theta^2 (-4d -2i\partial _l v^l-\square f)+\theta \sigma^l \bar \theta (-4i\partial _l n)+\\ 
+\theta^2 \bar \theta (\frac{1}{2}i\bar \sigma^l \partial_l )(-4\psi -2i\sigma^m \partial_m \bar \chi ) +\theta^2 \bar \theta^2 (-\square n) \\ \nonumber
D^2 X=-4m+\bar \theta (-4\bar \lambda  -2i\bar \sigma^l \partial_l \varphi )+\bar \theta ^2(-4d +2i\partial _l v^l-\square f)+\theta \sigma^l \bar \theta (4i\partial _l m)+\\ 
+\bar \theta^2 \theta (\frac{1}{2}i\sigma ^l \partial_l )(-4\bar \lambda - 2i\bar \sigma^m \partial_m \varphi ) +\theta^2 \bar \theta^2 (-\square m) 
\end{gather}
Note the signs in front of $\bar \sigma $ in (2.28), (2.29) which are different from the anticommuting case.  \\
We have further in the commuting case
\begin{gather}\nonumber
cX=\bar D^2 D^2 X=16d-8i\partial _l v^l+4\square f+\theta (8\square \varphi +16i\sigma^l \partial _l \bar \lambda )+ \\ \nonumber
+\theta^2 (16\square m)+\theta \sigma^l \bar \theta (16i\partial_l d+8\partial_l \partial_m v^m +4i\partial_l \square f)+ \\ 
+\theta^2 \bar \theta (8\square \bar \lambda -4i\bar \sigma^l \partial_l \square \varphi )+\theta^2 \bar \theta^2 (4\square d -2i\partial_l \square v^l + \square^2 f)
\end{gather}

\begin{gather}\nonumber
aX=D^2 \bar D^2 X=16d+8i\partial _l v^l+4\square f+\bar \theta (8\square
\bar \chi -16i\bar \sigma^l \partial _l \psi )+ \\ \nonumber
+\bar \theta^2 (16\square n)+\theta \sigma^l \bar \theta (-16i\partial_l d+8\partial_l \partial_m v^m -4i\partial_l \square f)+ \\ 
\bar\theta^2 \theta (8\square \psi +4i \sigma^l \partial_l \square \bar \chi )+\theta^2 \bar \theta^2 (4\square d +2i\partial_l \square v^l + \square^2 f)
\end{gather}
and finally obtain $T=-8\square +\frac{1}{2}(c+a)$ applied on $X$ as follows:

\begin{gather}\nonumber
TX= 16d-4\square f +\theta (-4\square \varphi +8i\sigma^l \partial _l \bar \lambda )+\bar \theta (-4\square \bar \chi  
-8i\bar \sigma^l \partial_l \psi )+\\ \nonumber
+\theta \sigma^l \bar \theta (8\partial_l \partial^m v_m -8\square v_l )+\theta^2 \bar \theta (-4\square \bar \lambda -2i\bar \sigma^l \partial_l \square \varphi )+\\ 
+\bar \theta^2 \theta  (-4\square \psi +2i\sigma^l \partial_l \square \bar \chi ) +\theta^2 \bar \theta^2 (-4\square d +\square^2 f)
\end{gather}
or in a form which makes transversality evident
\begin{gather}\nonumber
TX= 16d-4\square f +\theta (-4\square \varphi +8i\sigma^l \partial _l \bar \lambda )+\bar \theta (-4\square \bar \chi
+8i\bar \sigma^l \partial_l \psi )+\\ \nonumber
+\theta \sigma^l \bar \theta (8\partial_l \partial^m v_m -8\square v_l
)+\theta^2 \bar \theta (\frac{i}{2}\bar \sigma^l \partial_l )(-4\square \varphi +8i\sigma^l \partial _l \bar \lambda ) +\\ 
+\bar \theta^2 \theta (-\frac{i}{2}\sigma^l \partial_l ) (-4\square \bar 
\chi-8i\bar \sigma^l \partial _l \psi )+\theta^2 \bar \theta^2 (-4\square d +\square^2 f)
\end{gather}
Note that these formulas are very similar (but not identical) to the coresponding formulas deduced under the anticommuting spinor convention (see \cite{WB, Striv, C}). Again, they are connected by a change of sign of the Pauli matrix $\bar \sigma $.
We mention also the relations

\begin{gather}
\overline{D^2 X}=\bar D^2 \bar X ,\overline{\bar D^2 X}=D^2 \bar X \\
\overline{cX}=a\bar X ,\overline{aX}=c\bar X, \overline{TX}=\bar T \bar X
\end{gather}
These relations hold under the anticommuting convention too. 

\section{Superfunctions, superdistributions and the Hilbert-Krein structure associated to them}

Now we start defining a positive definite inner product on bona fide superfunctions, i.e. superfunctions of the form (1.1) with numerical (commuting) coefficints of the $\theta ,\bar \theta $-powers. These objects are definitely not those which appear in the standard supersymmetry literature sometimes under the ambiguous interchangeable name of "fields" or "functions". In fact the main point of this paper is that we make difference between supersymmetric functions and supersymmetric fields. The computations which we conduct under the supposition of commuting spinor components i.e. the commuting van der Waerden calculus of the previous section, are similar but not identical to the corresponding standard computations in supersymmetry. The main deviations from the standard case are the sign of the Pauli matrices $\bar \sigma $ and the complex conjugation which for instance produces $ -\bar \theta \bar \varphi $ in the commuting case (2.13) instead of $\bar \theta \bar \varphi $ in the anticommuting case (2.14) from the contribution $\theta \varphi $ in (1.1). \\
Let $d\rho (p)=\theta (-p_0 )\delta (p^2+m^2 )$ be the (tempered) Lorentz invariant measure concentrated on the closed backward mass hyperboloid in momentum space and let

\begin{gather}
D^+ (x)=\frac{1}{(2\pi )^2 }\int e^{ipx}d\rho (p)
\end{gather}

\begin{gather}
K_0 (z)=\delta^2 (\theta )\delta^2 (\bar \theta )D^+ (x)\\ 
K_c (z_1 ,z_2 )=P_c K_0 (z_1 -z_2 )\\
K_a (z_1 ,z_2 )=P_a K_0 (z_1 -z_2 )\\
K_T (z_1 ,z_2 )=-P_T K_0 (z_1 -z_2 )
\end{gather}
where $ \delta^2 (\theta^2 )=\theta^2, \delta^2 (\bar \theta^2 )=\bar \theta^2 $. The measure has to be concentrated on the backward instead of the forward light cone because of the specific choise of the Minkowski metric tensor adopted in this paper. 
Note the minus sign in front of $P_T $ which will play an important role in the sequel and the non-translation invariance of the last three kernels. Some factors of $\pi $ differ from the usual conventions. The plus superscript in (3.1) is standard. $ \theta $ in $ d \rho $ is the Heaviside function. \\
The material presented in the rest of this section looks similar to the one contained in \cite{C}. But we want to stress that here, in contrast to \cite{C}, we use the van der Waerden calculus under the commutative convention. This is definitely necessary if one wants to look at supersymmetric quantum fields as operator-valued (super)distributions. This is the standard of rigorous quantum field theory \cite{StreW} to which we subscribe.
For a general supersymmetric function $X$ let us identify $ X=\begin{pmatrix}X_c \\X_a \\X_T \end{pmatrix} $ where $X_c =X_1 =P_c X ,X_a =X_2 =P_a X,X_T =X_3 =P_T X, X=X_c +X_a +X_T $. We use the notation $X^T =(X_1 ,X_2 ,X_3 )$ for the transpose of $X$. We write $K_0 $ for $K_0 I_3 $ too where $I_3 $ is the 3x3-identity matrix. Let 

\begin{equation}\nonumber
\omega=\begin{pmatrix}1& 0 &0 \\0 &1& 0 \\0& 0 & -1 \end{pmatrix}
\end{equation}
Then we consider the inner products

\begin{equation}
<X,Y>=\int d^8z_1d^8z_2\bar X^T(z_1)K_0 (z_1-z_2) Y(z_2)
\end{equation}
and

\begin{equation}
(X,Y)=<X,\omega Y>=<\omega X,Y>
\end{equation}
Let us come back to $X=X_c +X_a +X_T,Y=Y_c +Y_a +Y_T $ in the identification $ X=\begin{pmatrix}X_c \\X_a \\X_T \end{pmatrix}, Y=\begin{pmatrix}Y_c \\Y_a \\Y_T \end{pmatrix} $. On the left hand side $X,Y$ are supersymmetric functions. We have in the above identification $\omega X=(P_c +P_a -P_T )X$ and after some elementary transformations we recognize $<.,.>$ and (.,.) to be

\begin{gather}\nonumber
<X,Y>=\int d^8z_1d^8z_2\bar X(z_1)^T (P_c +P_a +P_T )K_0 (z_1-z_2) Y(z_2)= \\
=\int d^8z_1d^8z_2\bar X(z_1)^T K_0 (z_1-z_2) Y(z_2)
\end{gather}
and

\begin{gather} \nonumber
(X,Y)=\int d^8z_1d^8z_2\bar X(z_1)^T (P_c +P_a -P_T )K_0 (z_1-z_2) Y(z_2)=\\ \nonumber
=\int d^8z_1d^8z_2\bar X(z_1)^T (K_c +K_a +K_T )(z_1,z_2) Y(z_2)= \\
=<X,(P_c +P_a -P_T )Y>
\end{gather}
respectively. Beside integration by parts in superspace we have used the fact that when applied to the kernel $K_0 (z_1-z_2 )$ the projection operators $P_i ,i=c,a,T $ can be freely moved from one variable to the other \cite{Striv}. One can prove on the lines of \cite{C} that $<.,.>$ is indefinite (not even semidefinite) whereas (.,.) is positive definite. This provides us with a standard Hilbert-Krein structure \cite{Stro}. Note the unexpected minus sign in front of $P_T $ in (3.9). By positivity we mean here non-negativity. Strict positivity will be achieved as usual by factorization of the zero vectors followed by completion. The positivity of $(.,.)$ is proved by computing it directly on the line of \cite{C} but there is a cheap way to understand this property. Indeed starting with the case $m>0$ we can decompose the problem on sectors. In the chiral sector for instance we realize that $(.,.)$ appears as the integral of a chiral function multiplied by its antichiral conjugate. It is then easy to see that it gives a positive contribution. In the transversal sector one has to use in addition the projection property of $P_T $ coupled with a geometric result in Minkowski space (see \cite{C} for the details of the similar computation using the anticommuting convention).\\
These arguments fail in the massless case but they  can be restored by restricting the space of test functions as explained below. One assumes that the supersymmetric functions $X,Y,\ldots $ are concentrated on the support of the measure $d\rho (p)$ in order to eliminate harmless zero vectors. In the massive case these are the only zero vectors. 
Some remarks are in order. 
First the problem with the d'Alembertian in the denominators of the formal projections. We want to look at them as projection operators in our Hilbert space. For the massive case $ m>0 $ their definition as Hilbert space projection operators makes no problems (the above restriction to the support of the measure $d\rho (p)$ means the on-shell condition). The Hilbert space segregated by the natural Krein structure introduced above decomposes into a direct sum of chiral, antichiral and transversal subspaces. The well defined projection operators are disjoint. But in the massless case the Hilbert space scalar product is badly defined because the explicite appearence of the formal, mathematically not well-defined, projections induced by the minus sign in front of $P_T $ (although the indefinite Krein space is perfectly defined because it contains no projection operators at all; otherwise stated, the projections sum up to one!). As a consequence the formal projections (in the massless case) cannot be generally realized as Hilbert space operators because of the d'Alembertian in the denominators. But there is a simple way out as follows (for more details in the case of the anticommuting convention see \cite{C}). Suppose the coefficient functions of the supersymmetric (test) functions satisfy mild restrictive conditions:

\begin{gather}
d(x)=\square D(x) \\
\bar \lambda (x)=\partial_l \Lambda \sigma^l \\
\psi (x)=\sigma^l \partial_l \Psi (x) \\
v_l (x)=\partial_l \rho +\omega, \partial_l \omega^l =0 
\end{gather}
where $D(x), \Lambda (x), \Psi (x),\rho (x), \omega (x) $ are arbitrary functions in $S$.
Just in order to have a name let us call them special supersymmetric functions. In particular the chiral, antichiral and transversal functions above are special supersymmetric.
Under these conditions the formal projections becomes well-defined. They segregate a d'Alembertian when applied to special supersymmetric functions. It follows that in the massless case the non-negative inner product space deduced from the Krein structure is well defined and $P_i ,i=c,a,T $ can be realized as projection operators (they are not yet disjoint because of the sector overlap). There is also in this case a simple way to see the positivity of $(.,.)$ without too much effort as in the massive case too. We leave the details to the reader. In order to produce the desirable Hilbert space we eliminate by factorization the overlap of the chiral/antichiral and trasversal sectors which turn out to consist exactly of the zero-vectors. The chiral, antichiral and transversal sectors become disjoint. Note that the conditions (3.10)-(3.13) are the supersymmetric counterpart of the divergence condition of the rigorous Gupta-Bleuler quantization \cite{Stro, C}.\\
Finally, from a technical point of view, we have now a Hilbert space of supersymmetric functions (with commuting spinor components) in which the formal operators (like for instance $D_\alpha ,\bar D_{\dot \alpha}, D^2,\bar D^2 ,a,c,T $) can be realized as bona fide Hilbert space operators with some useful properties. For instance it is easy to show that $\bar D^2 $ is the Hilbert space adjoint of $D^2 $. Even more, $ i\bar D_{\dot \alpha }$ (i being the unit imaginary) is the operator adjoint of $ iD_\alpha $, showing the analogy with the usual case of derivative $i\partial $ and Laplacian (or d'Alembertian). All this was made possible by giving up the anticommuting convention of spinor components. Under the anticommuting convention $D_\alpha ,\bar D_{\dot\alpha }  $ have no operator content. But we want to stress here that our commuting convention (together with the commuting van der Waerden calculus which was worked out in this paper) has to be applied for a rigurous supersymmetric quantum field theory only at the level of the test functions. As concernes supersymmetric fields, they certainly continue to carry bona fide spinors with anticommuting components. This point together with a canonical approach for the free chiral/antichiral fields was made clear in \cite{C2}.\\
The main issue of this paper in conjunction with \cite{C} was to present a rigorous framework for quantum supersymmetric fields as operator-valued superdistributions with an accent on both test superfunctions and  Hilbert-Krein structure which is naturaly carried by them. For this purpose we use a commuting variant of the van der Waerden calculus elaborated in Section 2 of this paper. The results should be useful in any formulation of supersymmetric quantum field theory in which positivity, and as such unitarity, is manifest. \\
Finally note that the extension to the $ N>1$ supersymmetry is possible. The prerequisites of this generalization are contained in \cite{GGRS}, p. 121-127.


\begin{thebibliography}{99}

\bibitem {C} F. Constantinescu, J. Phys. A: Math. Gen. 38(2005), 1385
\bibitem {Stro} F. Strocchi, A.S. Wightman, Journal Math. Phys. 12(1974), 2198, F. Strocchi, Selected Topics on the General Properties of Quantum Field Theory, World Scientific, 1993
\bibitem {WB} J. Wess, J.Bagger, Supersymmetry and Supergravity, 2nd edition, Princeton University Press, 1992
\bibitem {StreW} R.F. Streater, A.S. Wightman, PCT, Spin and Statistics and All That, Benjamin, 1964
\bibitem {Striv} P.P. Strivastava, Supersymmetry, Superfields and Supergravity: An Introduction, IOP Publishing, Adam Hilger, Bristol, 1986
\bibitem {Wei} S. Weinberg, The Quantum Theory of Fields, vol III, Cambridge University Press, 1996
\bibitem {vdW} B.L. van de Waerden, Die gruppentheoretische Methode in der Quantenmechanik, Springer Verlag, 1932
\bibitem {PR} R. Penrose, W. Rindler, Spinors and Space-Time, Vol I, Cambridge University Press, Cambridge 1986
\bibitem {C2} F. Constantinescu, Lett. Math. Phys. 62(2002), 111
\bibitem {GGRS} S.J. Gates Jr., M.T. Grisaru, M. Rocec, W. Siegel, Superspace, Benjamin, 1983

\end{thebibliography}
\end{document}